\documentclass[nofootinbib,preprint,preprintnumbers,a4paper,11pt]{revtex4}
\usepackage{epsfig}
\usepackage{footnote}
\usepackage{ulem}
\usepackage{color}
\usepackage{array}
\usepackage{amssymb}
\usepackage{amsmath}
\usepackage{graphicx}
\usepackage{subfigure}
\usepackage{longtable}
\usepackage{verbatim}
\usepackage{amsfonts}
\usepackage{hyperref}
\usepackage{multirow}
\usepackage{cancel}

\begin{document}
\title{Evidence of $A_{CP}(D^0\to \pi^+\pi^-)$ implies \\ observable $CP$ asymmetry in the $D^0\to \pi^0\pi^0$ decay}
\author{Di Wang}
\email{wangdi@hunnu.edu.cn}
\affiliation{Department of Physics, Hunan Normal University, and Key Laboratory of Low-Dimensional Quantum Structures and Quantum Control of Ministry of Education, Changsha 410081, China}

\begin{abstract}
Inspired by the recent measurement of $CP$ asymmetry in the individual mode on LHCb, we study $CP$ asymmetry of the $D\to \pi\pi$ system in the isospin and topological analysis.
The ratio between penguin and tree amplitudes $P/(T+C)$ in the $D\to \pi\pi$ system is found to be greater than two in most values of the relative strong phase.
And $D^0\to \pi^0\pi^0$ is a potential mode to reveal the $CP$ asymmetry of the order of $10^{-3}$, which would be observed by Belle II in the future.
The large $CP$ asymmetry in the $D\to \pi\pi$ system might be understood in the $t$-channel final-state interaction.

\end{abstract}

\maketitle

\section{Introduction}

$CP$ asymmetry in $D$ meson decay, which is defined as
\begin{align}
A_{CP}(D\to f) \equiv \frac{\Gamma(D\to f)-\Gamma(\overline D\to \overline f)}{\Gamma(D\to f)+\Gamma(\overline D\to \overline f)},
\end{align}
provides a window to test the Standard Model (SM) and search for new physics (NP) in the up-type quark weak decay in hadrons.
The LHCb collaboration observed $CP$ asymmetry of charm decay in 2019 \cite{LHCb:2019hro},
\begin{align}\label{z1}
  \Delta A_{CP} &\equiv A_{CP}(D^0\to K^+K^-)-A_{CP}(D^0\to \pi^+\pi^-)=(-1.54\pm0.29)\times 10^{-3}.
\end{align}
After that, many experimental efforts are devoted to the measurement of $CP$ asymmetry and mixing parameters in charm system \cite{LHCb:2022gnc,LHCb:2021dcr,LHCb:2021ykz,LHCb:2021vmn,LHCb:2021rdn,LHCb:2021rou,Belle:2019xha}.
In theoretical aspect, there are two controversial viewpoints for the observed $CP$ asymmetry difference in literature, regarding it as signal of new physics \cite{Chala:2019fdb,Dery:2019ysp,Calibbi:2019bay,Buras:2021rdg}, or the non-perturbative QCD enhancements to penguin \cite{Li:2012cfa,Cheng:2012xb,Soni:2019xko,Bediaga:2022sxw,Wang:2021rhd,Schacht:2021jaz,Li:2019hho,Cheng:2019ggx,Grossman:2019xcj}.
It attributes to the large ambiguities in evaluating penguin topologies and the absence of more information given by experiments.

Very recently, the LHCb collaboration reported the first evidence of non-vanishing $CP$ asymmetry in the individual decay of $D^0\to \pi^+\pi^-$ by measuring $\Delta A_{CP}$ and $CP$ asymmetry in the $D^0\to K^+K^-$ decay \cite{pipi}.
The $CP$ asymmetries in the $D^0\to K^+K^-$ and $D^0\to \pi^+\pi^-$ decays are given by
\begin{align}
  A_{CP}(D^0\to K^+K^-) &= (0.77\pm 0.57)\times 10^{-3}, \\  A_{CP}(D^0\to \pi^+\pi^-) &= (2.32\pm 0.61)\times 10^{-3}.
\end{align}
$CP$ asymmetry in the individual mode is more significant compared to the difference between two decay modes because it allows us to extract more knowledge of non-perturbative QCD.
The newest data indicates a very large $U$-spin breaking in the $D^0\to K^+K^-$ and $D^0\to \pi^+\pi^-$ modes, which is beyond the naive expectations of $\varepsilon \sim m_s/\Lambda_{\rm QCD}\sim 30\%$ \cite{Schacht:2022kuj}.

In this work, we analyze the implications of the new measurement of $CP$ asymmetry in the $D^0\to \pi^+\pi^-$ decay.
By applying the isospin and topological analysis, we show the ratio between penguin and tree amplitudes in the $D\to \pi\pi$ system is greater than $2$ in most values of strong phase.
And $CP$ asymmetry in the $D^0\to \pi^0\pi^0$ decay could reach to be $\mathcal{O}(10^{-3})$, which is available on Belle II in the future.
The large $CP$ asymmetry in the $D\to \pi\pi$ system might be understood in the $t$-channel final-state interaction (FSI) \cite{Han:2021azw,Jiang:2018oak,Li:2020qrh,Chen:2020eyu,Chen:2002jr,Han:2021gkl,Yu:2017zst,Lu:2005mx,Cheng:2004ru,Locher:1993cc,Dai:1999cs,Li:1996cj,Ablikim:2002ep,Li:2002pj}.

This paper is organized as follows.
In Sec.~\ref{topox}, we analyze $CP$ asymmetry in the $D\to \pi\pi$ system in the isospin and topological analysis.
In Sec.~\ref{fsi1}, we try to explain the large $CP$ asymmetry in charm in the final state interaction.
And Sec.~\ref{con} is a short summary.

\section{Isospin and topological analysis}\label{topox}

In the $D\to \pi\pi$ system, $(D^0,D^+)$ form an isospin doublet, $(\pi^+,\pi^0,\pi^-)$ form an isospin triplet.
Isospin decompositions of the $D^0\to \pi^+\pi^-$, $D^0\to \pi^0\pi^0$ and $D^+\to \pi^+\pi^0$ modes are
\begin{align}\label{iso1}
\mathcal{A}(D^0\to \pi^+\pi^-) &= \frac{1}{2\sqrt{3}}\mathcal{A}_{3/2}+\frac{1}{\sqrt{6}}\mathcal{A}_{1/2}, \\ \label{iso2}
\mathcal{A}(D^0\to \pi^0\pi^0) &= \frac{1}{\sqrt{6}}\mathcal{A}_{3/2}-\frac{1}{2\sqrt{3}}\mathcal{A}_{1/2},  \\ \label{iso3}
\mathcal{A}(D^+\to \pi^+\pi^0) & =\frac{\sqrt{6}}{4}\mathcal{A}_{3/2},
\end{align}
in which $\mathcal{A}_{3/2}$ and $\mathcal{A}_{1/2}$ are the amplitudes with $\Delta I = 3/2$ and $\Delta I = 1/2$, respectively.
Topological decompositions of the $D\to \pi\pi$ modes can be expressed as
\begin{align}\label{topo1}
\mathcal{A}(D^0\to \pi^+\pi^-) &= \lambda_d\,(T+E+\mathcal{P}_{\rm b})-\lambda_b\,\mathcal{P}, \\ \label{topo2}
\mathcal{A}(D^0\to \pi^0\pi^0) &= \frac{1}{\sqrt{2}}\lambda_d\,(C-E-\mathcal{P}_{\rm b})+\frac{1}{\sqrt{2}}\lambda_b\,\mathcal{P},  \\ \label{topo3}
\mathcal{A}(D^+\to \pi^+\pi^0) & =\frac{1}{\sqrt{2}}\lambda_d\,(T+C),
\end{align}
where $\lambda_d=V^*_{cd}\,V_{ud}$ and $\lambda_b=V^*_{cb}\,V_{ub}$.
The contributions from the penguin operators $O_{3-6}$ and the chromomagnetic penguin operator $O_{8g}$ are neglected in Eqs.~\eqref{topo1}$\sim$\eqref{topo3}.
$T$, $C$ and $E$ denote the tree amplitudes and $\mathcal{P}$ and $\mathcal{P}_{\rm b}$ denote the penguin amplitudes.
$\mathcal{P}_{\rm b}$ is the difference between $\mathcal{P}_d$ and $\mathcal{P}_s$, $\mathcal{P}_{\rm b} = \mathcal{P}_d - \mathcal{P}_s$, and $\mathcal{P}=\mathcal{P}_s$. $\mathcal{P}_d$ and  $\mathcal{P}_s$ are the topologies with $d$ and $s$ in the quark loop respectively.
In literature such as \cite{Muller:2015lua}, $\mathcal{P}$ is written as penguin plus penguin annihilation diagrams, and $\mathcal{P}_{\rm break}$ is defined as $\mathcal{P}_{\rm break} = \mathcal{P}_s - \mathcal{P}_d = -\mathcal{P}_{\rm b} $.
In order to math the isospin amplitudes, the quark compositions of $D$ and $\pi$ mesons are defined as $D^0 = -c\overline u$, $D^+ = c\overline d$, $\pi^+ = u\overline d$, $\pi^0 = \frac{1}{\sqrt{2}}\,(d\overline d - u\overline u)$ and $\pi^- = -d\overline u$ in Eqs.~\eqref{topo1}$\sim$\eqref{topo3}.
By comparing Eqs.~\eqref{topo1}$\sim$\eqref{topo3} with Eqs.~\eqref{iso1}$\sim$\eqref{iso3}, the relations between isospin amplitudes and topological amplitudes are found to be
\begin{align}\label{3/2}
\mathcal{A}_{3/2} &= \frac{2\sqrt{3}}{3}\lambda_d(T+C), \\ \label{1/2}
\mathcal{A}_{1/2} & = \sqrt{6}\lambda_d(E+\mathcal{P}_{\rm b}+\frac{2}{3}T-\frac{1}{3}C)-\sqrt{6}\lambda_b \mathcal{P}.
\end{align}
Eqs.~\eqref{3/2} and \eqref{1/2} can also be derived from the effective Hamiltonian of charm decay by analyzing the isospin structure of tree and penguin operators, see literature such as Ref.~\cite{Wang:2020gmn} for details.

In the SM, $\lambda_b$ is much smaller than $\lambda_d$, $\lambda_b/\lambda_d \sim \mathcal{O}(10^{-4})$~\cite{Workman:2022ynf}.
The last term in Eq.~\eqref{1/2} can be neglected safely in the branching fractions.
We define an approximate $\Delta I =1/2$ amplitude without the $\lambda_b\, P$ term as
\begin{align}
\mathcal{A}^\prime_{1/2} = \sqrt{6}\,\lambda_d\,(E+\mathcal{P}_{\rm b}+\frac{2}{3}T-\frac{1}{3}C).
\end{align}
$\mathcal{A}^\prime_{1/2}$ can be written as
\begin{align}
\mathcal{A}^\prime_{1/2} = \mathcal{A}^{\prime s}_{1/2}\,e^{i\,\delta_{I}},
\end{align}
with magnitude $\mathcal{A}^{\prime s}_{1/2}$ and relative strong phase $\delta_{I} =\delta_{1/2} - \delta_{3/2}$. The strong phase of $\mathcal{A}_{3/2}$ is usually set to be zero.
The isospin amplitudes $\mathcal{A}_{3/2}$ and $\mathcal{A}^\prime_{1/2}$ can be extracted from the branching fractions of three $D\to \pi\pi$ modes which are given by \cite{Workman:2022ynf}
\begin{align}
  \mathcal{B}r(D^0\to \pi^+\pi^-) &= (1.454\pm 0.024)\times 10^{-3}, \qquad \mathcal{B}r(D^0\to \pi^0\pi^0) = (0.826\pm 0.025)\times 10^{-3},  \nonumber\\
    \mathcal{B}r(D^+\to \pi^+\pi^0)& =(1.247\pm 0.033)\times 10^{-3}.
\end{align}
The partial decay width $\Gamma$ is parameterized to be
\begin{align}
\Gamma(D\to \pi\pi) = \frac{|P_c|}{8\pi m^2_{D}}|\mathcal{A}(D\to \pi\pi)|^2,
\end{align}
in which $P_c$ is the $c.m.$ momentum in the rest frame of $D$ meson.
The hadronic parameters $\mathcal{A}_{3/2}$, $\mathcal{A}^{\prime s}_{1/2}$ and $\delta_{I}$ are extracted to be
\begin{align}\label{re}
& \mathcal{A}_{3/2} = (0.447\pm 0.006)\times 10^{-6}\,{\rm GeV},\qquad \mathcal{A}^{\prime s}_{1/2} = (1.090\pm 0.009)\times 10^{-6}\,{\rm GeV}, \nonumber\\
&\delta^n_{I} = (-86.66\pm 1.31)^\circ,\qquad\qquad \delta^p_{I} = (86.66\pm 1.31)^\circ.
\end{align}
There are two solutions for the strong phase $\delta_I$.
Superscripts $n$ and $p$ are used to distinguish the negative and positive solutions.
Eq.~\eqref{re} is consistent with the result given by Ref.~\cite{Franco:2012ck}.

To analyze $CP$ asymmetries in the $D\to \pi\pi$ modes, we parameterize the penguin amplitude $\mathcal{P}$ as
\begin{align}
\lambda_b\,\mathcal{P} = |\lambda_b|\,P\,e^{i\,(\delta_p - \gamma)},
\end{align}
in which $P$ and $\delta_p$ are the magnitude and strong phase (with respect to $A_{3/2}$) of penguin amplitude respectively.
$\gamma$ is phase parameter of the CKM matrix, known as $\phi_3$ in the unitarity triangle.
In the SM, $\gamma$ is fitted to be $1.144\pm 0.027$ \cite{Workman:2022ynf}.
The weak phase of $\lambda_d$ is negligible compared to $\gamma$.
With the isospin and topological amplitudes, $CP$ asymmetries of the $D^0\to \pi^+\pi^-$ and $D^0\to \pi^0\pi^0$ decays are derived to be
\begin{align}\label{cp1}
A_{CP}(D^0\to \pi^+\pi^-) = 4\sqrt{3}\,\frac{|\lambda_b|}{\lambda_d}\,\frac{ \,P\sin\gamma\,\left(\sqrt{2}\,\mathcal{A}_{1/2}^{\prime s}\sin(\delta_I-\delta_p)-\mathcal{A}_{3/2}\sin\delta_p\right) }{2(\mathcal{A}_{1/2}^{\prime s})^2+\mathcal{A}_{3/2}^2+2\sqrt{2}\,\mathcal{A}_{1/2}^{\prime s}\mathcal{A}_{3/2}\cos\delta_I},
\end{align}
\begin{align}\label{cp2}
A_{CP}(D^0\to \pi^0\pi^0) = 2\sqrt{3}\,\frac{|\lambda_b|}{\lambda_d}\,\frac{ \,P\sin\gamma\,\left(\sqrt{2}\,\mathcal{A}_{1/2}^{\prime s}\sin(\delta_I-\delta_p)+2\,\mathcal{A}_{3/2}\sin\delta_p\right) }{(\mathcal{A}_{1/2}^{\prime s})^2+2\,\mathcal{A}_{3/2}^2-2\sqrt{2}\,\mathcal{A}_{1/2}^{\prime s}\mathcal{A}_{3/2}\cos\delta_I}.
\end{align}
In Eqs.~\eqref{cp1} and \eqref{cp2}, the first term in numerator is the $CP$ asymmetry in $\mathcal{A}_{1/2}$ and the second term is the interference between $\mathcal{A}_{1/2}$ and $\mathcal{A}_{3/2}$.

\begin{figure}
  \centering
  \includegraphics[width=0.45\textwidth]{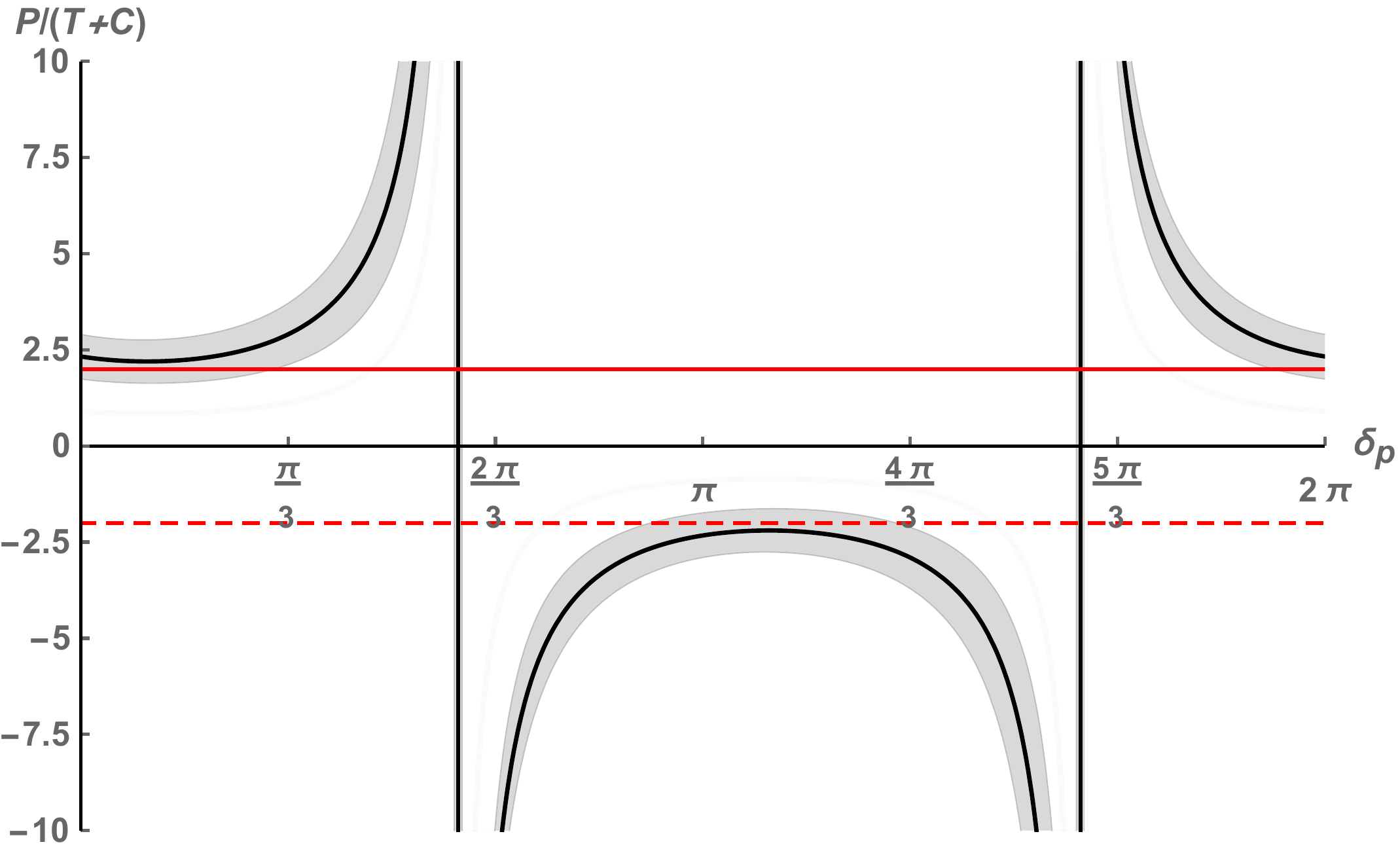}
  \qquad\quad
  \includegraphics[width=0.45\textwidth]{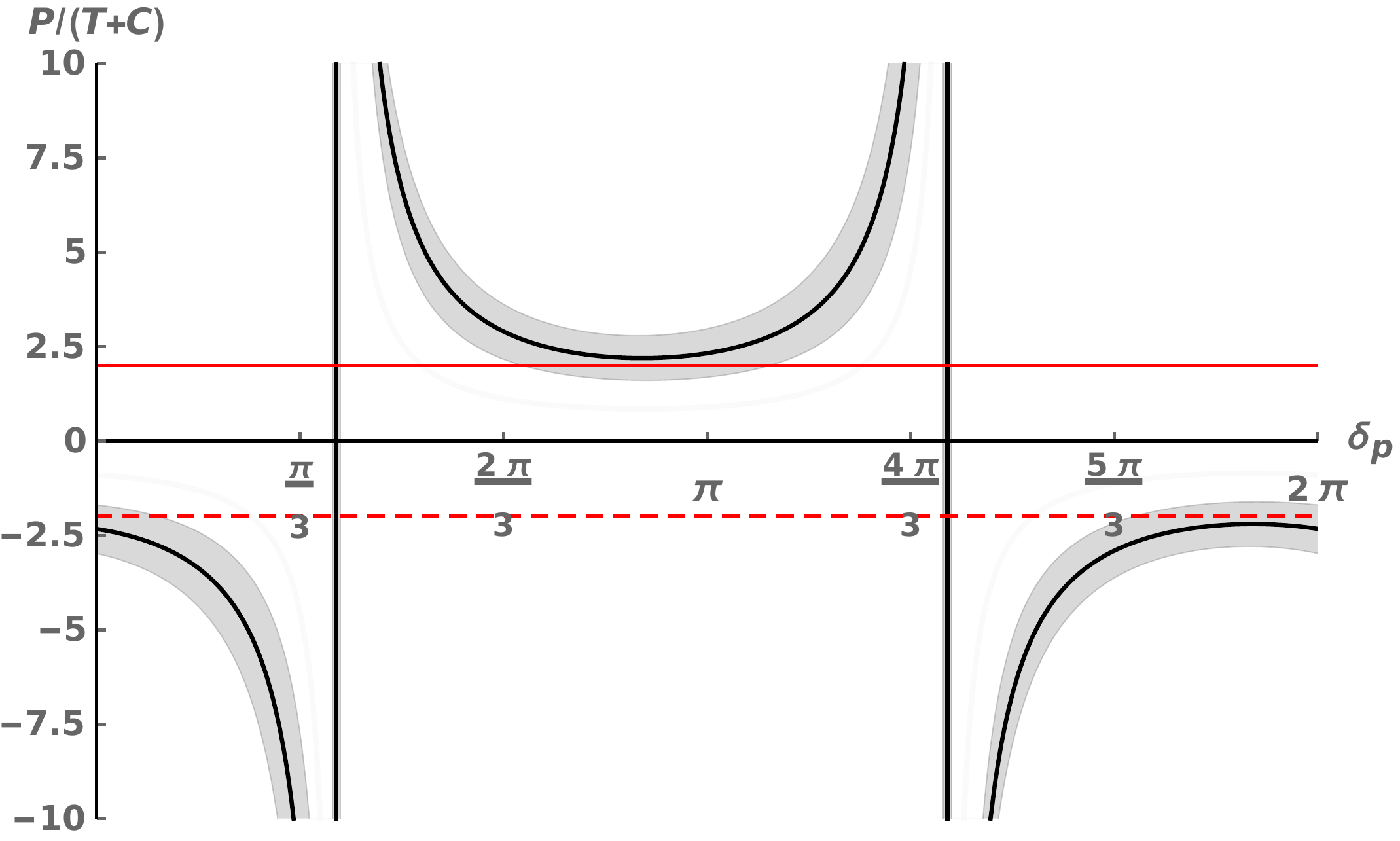}
\caption{Ratio of penguin and tree amplitudes $P/(T+C)$ dependent on the strong phase $\delta_p$ in the cases of negative (left) and positive (right) strong phase $\delta_I$.
The gray bands represent the uncertainties, and the blue solid and dashed lines are $P/(T+C)=2$ and $-2$ respectively. }
\label{p}
\end{figure}
There are two non-determined parameters in Eqs.~\eqref{cp1} and \eqref{cp2}, $P$ and $\delta_p$.
If the scenario of no new physics effect is assumed, we can solve $P$ as a function of $\delta_p$ according to the experiment result of $A_{CP}(D^0\to \pi^+\pi^-)$.
The ratio between penguin and tree amplitudes $P/(T+C)$ dependent on $\delta_p$ is plotted in Fig.~\ref{p}.
One can find $P/(T+C)$ is greater than $2$ in most values of $\delta_P$ in both negative and positive $\delta_I$.
It suggests the penguin topology is enhanced by non-perturbative QCD in the $D\to \pi\pi$ system.
More generally, it is possible that the large penguin topologies exist in other singly Cabibbo-suppressed charmed hadron decay modes, leading to observable $CP$ asymmetries.
Topology $\mathcal{P}_{\rm b}$ is comparable to the tree amplitudes, contributing to a large $SU(3)_F$ breaking effect in the singly Cabibbo-suppressed charm decays and affecting the branching fractions.
Thus $\mathcal{P}_{\rm b}$ cannot be neglected in the global fit of the $D$ meson or baryon non-leptonic weak decays.

\begin{figure}
  \centering
  \includegraphics[width=0.45\textwidth]{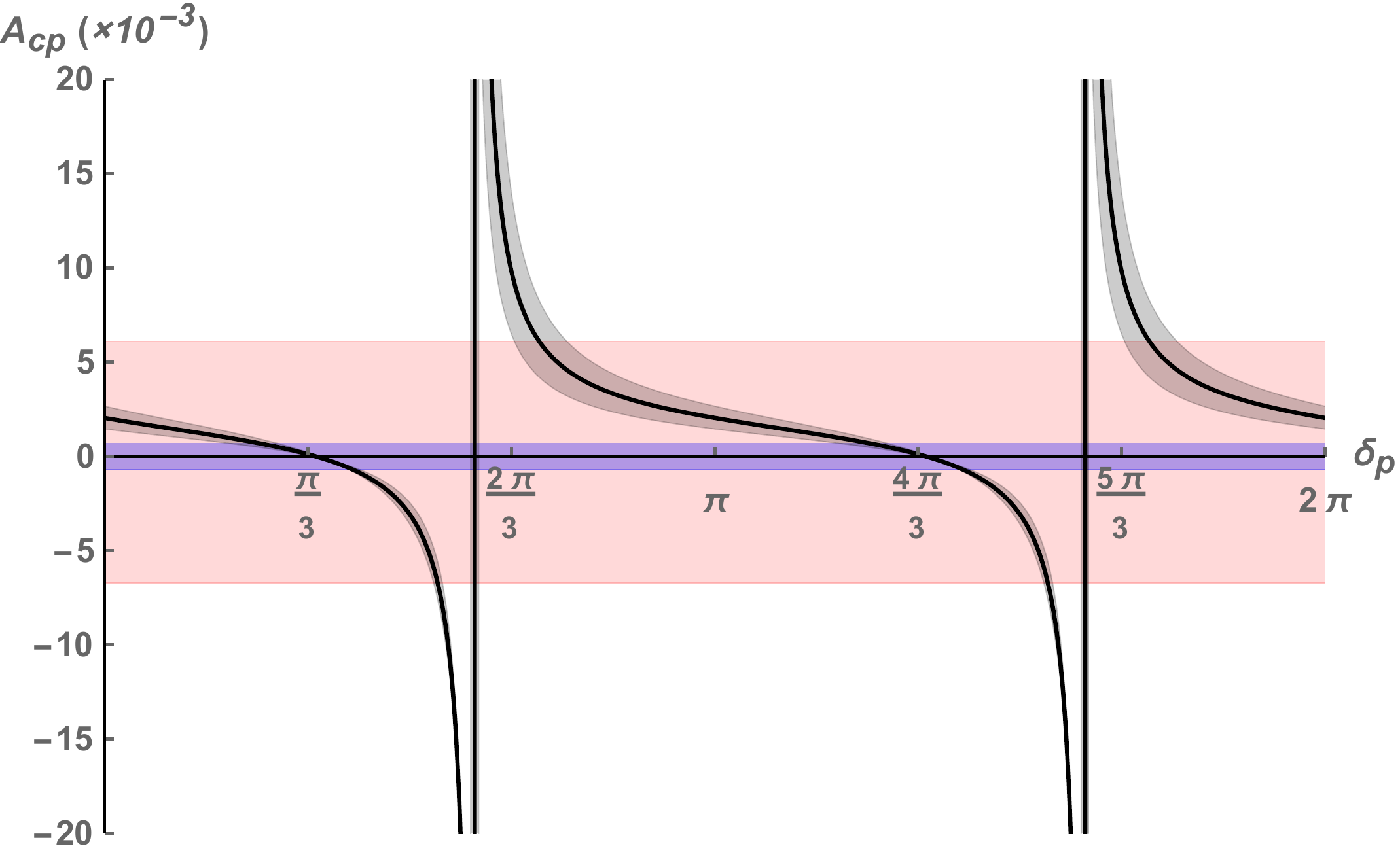}
  \qquad\quad
  \includegraphics[width=0.45\textwidth]{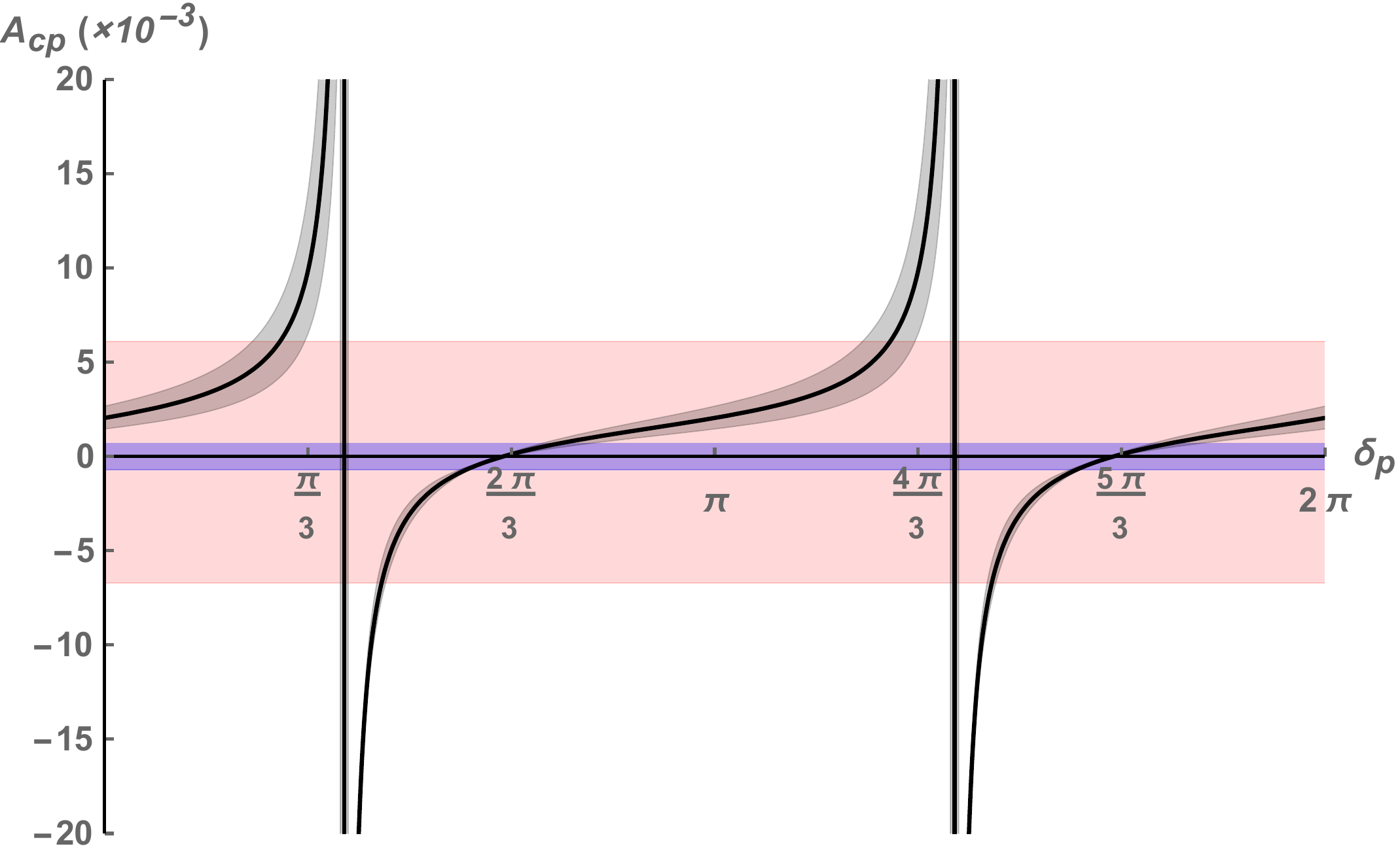}
\caption{$CP$ asymmetry in the $D^0\to \pi^0\pi^0$ decay dependent on $\delta_p$ in the cases of negative (left) and positive (right) $\delta_I$.
The horizontal pink shadow is $1\sigma$ experimental limitation to date~\cite{Belle:2014evd,Workman:2022ynf}.
The blue shadow is the expected statistical uncertainties on Belle II at $50{\rm ab}^{-1}$ data set \cite{Belle-II:2022cgf}.}
\label{cp}
\end{figure}
With the function of $P(\delta_p)$, we get the function of $A_{CP}(D^0\to \pi^0\pi^0)$ dependent on $\delta_p$, which is plotted in Fig.~\ref{cp}. $A_{CP}(D^0\to \pi^0\pi^0)$ is expected to be $\mathcal{O}(10^{-3})$  at most values of $\delta_p$, which is available on Belle II at $50{\rm ab}^{-1}$ data set.
At some particular values of $\delta_p$, $A_{CP}(D^0\to \pi^0\pi^0)$ could reach to be $\mathcal{O}(10^{-2})$.
So $CP$ asymmetry in the $D^0\to \pi^0\pi^0$ decay might be the next observed $CP$ asymmetry in charm sector.
On the other hand, if $CP$ asymmetries of the $D^0\to \pi^+\pi^-$ and $D^0\to \pi^0\pi^0$ decays are well determined by experiments, the magnitude and strong phase of penguin $\mathcal{P}$ can be extracted without model calculations.
It will deepen the understanding of non-perturbative QCD in charm scale.

\begin{figure}
  \centering
  \includegraphics[width=0.45\textwidth]{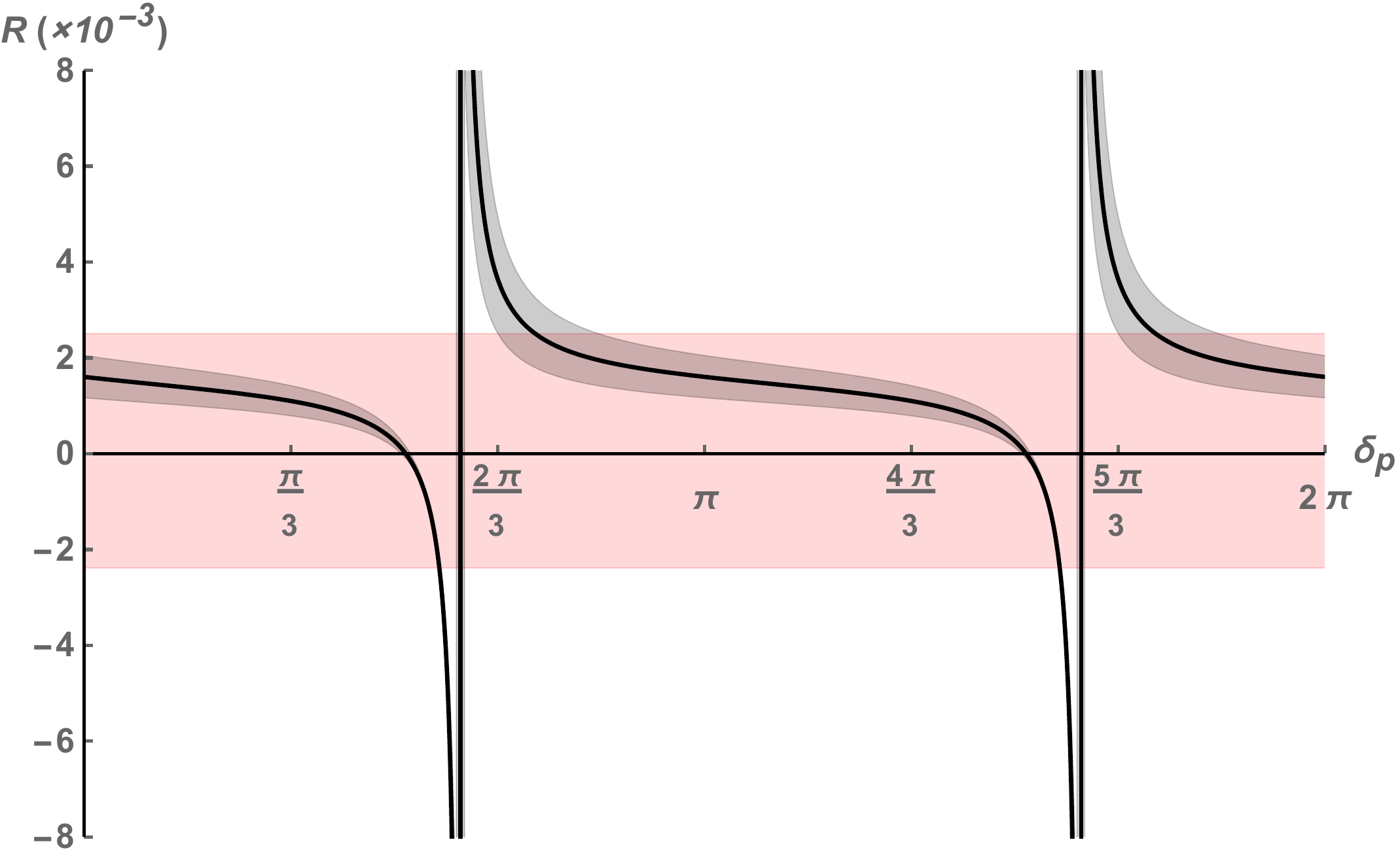}
  \qquad\quad
  \includegraphics[width=0.45\textwidth]{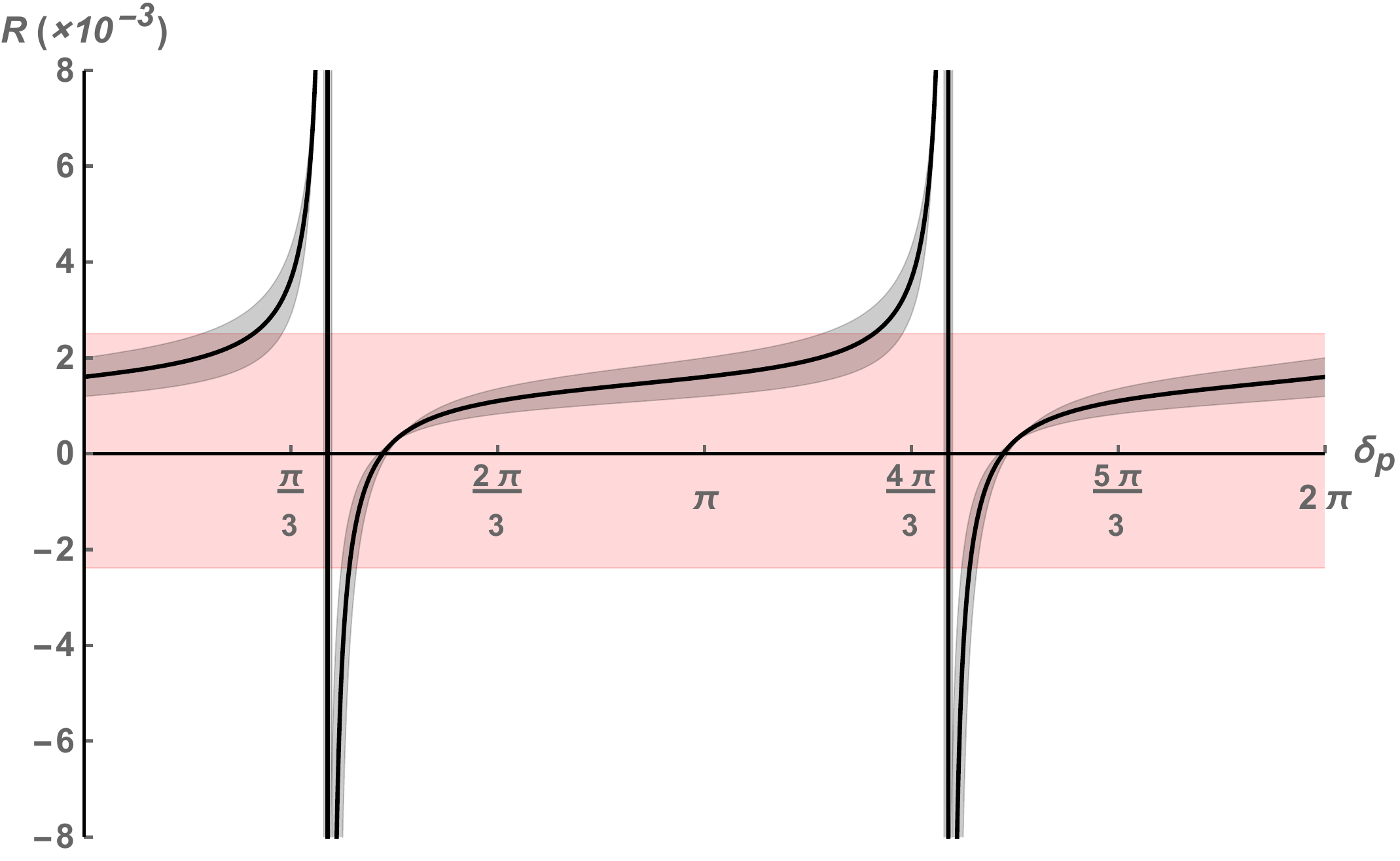}
\caption{Ratio $R$ dependent on $\delta_p$ in the cases of negative (left) and positive (right) $\delta_I$, in which the horizontal pink shadow is $1\sigma$ experimental limitation to date taken from HFLAV~\cite{HFLAV:2022pwe}. }
\label{R}
\end{figure}
In experiments, the ratio $R$ is used to test new physics in the $\Delta I = 3/2$ amplitude, which is defined by \cite{Grossman:2012eb,Belle:2017tho}
\begin{align}
R=\frac{|\mathcal{A}(\pi^+\pi^-)|^2-|\overline{\mathcal{A}}(\pi^+\pi^-)|^2
+|\mathcal{A}(\pi^0\pi^0)|^2-|\overline{\mathcal{A}}(\pi^0\pi^0)|^2
-\frac{2}{3}(|\mathcal{A}(\pi^+\pi^0)|^2-|\overline{\mathcal{A}}(\pi^+\pi^0)|^2)}
{|\mathcal{A}(\pi^+\pi^-)|^2+|\overline{\mathcal{A}}(\pi^+\pi^-)|^2
+|\mathcal{A}(\pi^0\pi^0)|^2+|\overline{\mathcal{A}}(\pi^0\pi^0)|^2
+\frac{2}{3}(|\mathcal{A}(\pi^+\pi^0)|^2+|\overline{\mathcal{A}}(\pi^+\pi^0)|^2)}.
\end{align}
With the isospin and topological amplitudes of $D\to \pi\pi$ modes, ratio $R$ is derived to be
\begin{align}\label{rr}
 R= 6\sqrt{6}\,\frac{|\lambda_b|}{\lambda_d}\,\frac{ \,\mathcal{A}_{1/2}^{\prime s}\,P\,\sin\gamma\,\sin(\delta_I-\delta_p) }{3\,(\mathcal{A}_{1/2}^{\prime s})^2+11\,\mathcal{A}_{3/2}^2}
\end{align}
in the SM. According to Eq.~\eqref{rr}, ratio $R$ is determined by $CP$ asymmetry in the $\Delta I = 1/2$ amplitude.
The dependence of ratio $R$ on $\delta_p$ is plotted in Fig.~\ref{R}.
It is found that $R \sim \mathcal{O}(10^{-3})$ in the most values of $\delta_p$.

\section{Estimation in the final-state interaction}\label{fsi1}

In the $D\to \pi\pi$ system, the penguin contribution is $\lambda_d\mathcal{P}_d+\lambda_s\mathcal{P}_s$.
Considering all the tree amplitudes are proportional to $\lambda_d$ in the $D\to \pi\pi$ modes, we write
penguin amplitudes as $\lambda_d\mathcal{P}_d+\lambda_s\mathcal{P}_s = \lambda_d\mathcal{P}_d-(\lambda_d+\lambda_b)\mathcal{P}_s = \lambda_d\mathcal{P}_{\rm b}-\lambda_b\mathcal{P}_s$ in Eqs.~\eqref{topo1}$\sim$ \eqref{topo3}. The $CP$ asymmetry is induced by the interference between $\lambda_b\mathcal{P}_s$ with other decay amplitudes.
In this section, we discuss how large the penguin amplitude $\mathcal{P}_s$ could be in the final-state interaction.

\begin{figure}
  \centering
  \includegraphics[width=0.70\textwidth]{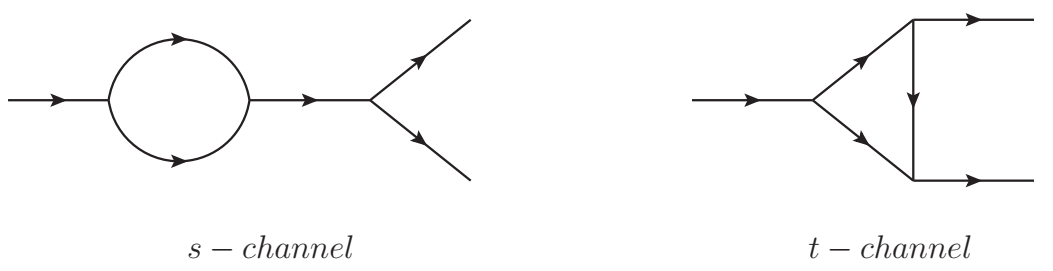}
\caption{Final-state interaction effects: $s$-channel (resonance) and $t$-channel (rescattering) contributions to one particle exchange. }
\label{st}
\end{figure}

For the two-body heavy meson weak decay, the FSI effect can be modeled as  exchange of one particle between two particles generated from the short-distance tree emitted process.
There are $s$-channel and $t$-channel contributions in the final state interaction, which are depicted in Fig.~\ref{st}.
In the $s$-channel contribution, the resonance state in the $D\to \pi\pi$ decay has the quantum number $J^{PC}=0^{++}$ derived from the final states. Ref.~\cite{Soni:2019xko} suggests that $f_0(1710)$ playing an important role in enhancing the penguin amplitude.
However, the $CP$ asymmetry ratio between $D^0\to \pi^+\pi^-$ and $D^0\to K^+K^-$ modes is expected to be $-1.06$, which is different from the experimental result.
It is attributed to the suppression by the ratio of branching fraction, $\mathcal{B}r(f_0(1710)\to \pi\pi)/\mathcal{B}r(f_0(1710)\to KK)\approx 0.4$.
If the $t$-channel contribution is included, the large $CP$ asymmetry in the $D\to\pi\pi$ might be understood in the FSI.

\begin{figure}
\centering
\includegraphics[width=0.97\textwidth]{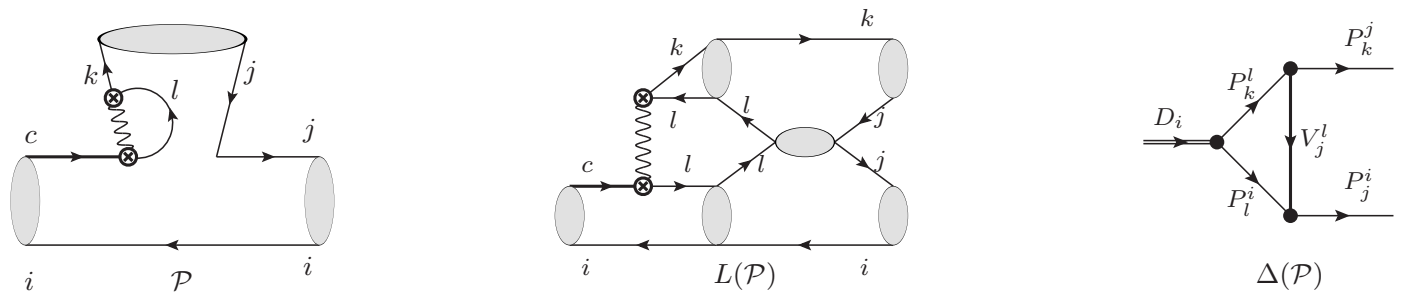}
\caption{Penguin diagram and re-scattering contribution to the penguin diagram. }
\label{topo}
\end{figure}

An approach to relate topological diagram to the re-scattering triangle diagram was proposed in Ref.~\cite{Wang:2021rhd}.
The re-scattering contribution in the penguin amplitude can be described by Fig.~\ref{topo}.
In the topological diagram $\mathcal{P}$ in Fig.~\ref{topo}, indices $i$, $j$, $k$ and $l$ present the light quark $u$, $d$ or $s$ in the charm decay. $\mathcal{P}$ diagram can be obtained by twisting quark lines from a short-distance $T$ diagram, as shown in the second diagram, $L(\mathcal{P})$, in Fig.~\ref{topo}.
$L(\mathcal{P})$ forms a triangle diagram  at hadron level, which is the third diagram in Fig.~\ref{topo}.
In the triangle diagram, the vertex in the left is a weak vertex and the other two vertexes are strong vertexes.
The quark loop in penguin can be understood as the light quark $q_l$ exited from the weak vertex goes through three propagators in the triangle diagram then returns to the weak vertex.

\begin{figure}
  \centering
  \includegraphics[width=0.65\textwidth]{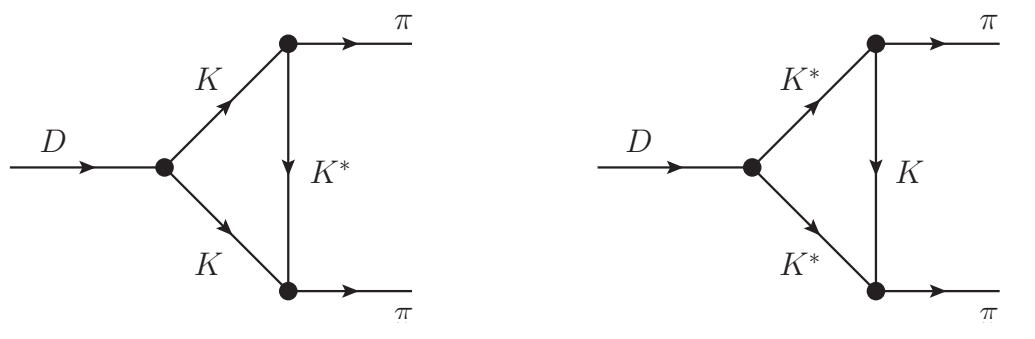}
\caption{Triangle diagrams contributing to $CP$ asymmetries in the $D\to \pi\pi$ system. }
\label{tri}
\end{figure}
There are two triangle diagrams contributing to the penguin diagram $\mathcal{P}_s$ shown in Fig.~\ref{tri}.
The left diagram presents $KK\to \pi\pi$ scattering and the right diagram presents $K^*K^*\to \pi\pi$ scattering.
Compared to Ref.~\cite{Schacht:2022kuj}, the $K^*K^*\to \pi\pi$ scattering is included in the final state interaction.
Since the weak vertex of triangle diagram is a short-distance $T$ diagram, the factorization approach is used to estimate the weak decay amplitude.
It is parameterized as the decay constant of the emitted meson and the transition form factor of another meson.
For the amplitude of total triangle diagram, there are several calculational methods, in which the treatment of hadronic loop integration is different \cite{Lu:2005mx,Cheng:2004ru,Locher:1993cc,Dai:1999cs,Li:1996cj,Ablikim:2002ep,Li:2002pj}.
In this work, the intermediate states are treated to be on their mass shell and then the optical theorem and Cutkosky cutting rule are used to calculate the absorptive part. The calculation details can be found in Ref.~\cite{Ablikim:2002ep}.

Considering the exchanged meson being generally off-shell, a form factor is introduced as $F(t) = (\Lambda^2-m_t^2)/(\Lambda^2-t)$ to compensate the off-shell effect \cite{Gortchakov:1995im}, where $t$ and $m_t$ are the momentum square and mass of the exchanged meson respectively.
$F(t)$ is normalized to unity at the on-shell situation $t=p^2_{t}=m_t^2$.
The cutoff $\Lambda$ is parameterized as $\Lambda = m_{t} + \eta\Lambda_{\rm QCD}$
with $\Lambda_{\rm QCD} = 330\,{\rm MeV}$ for the charm decay.
The parameter $\eta$ cannot be calculated from the QCD method.

\begin{figure}
  \centering
  \includegraphics[width=0.50\textwidth]{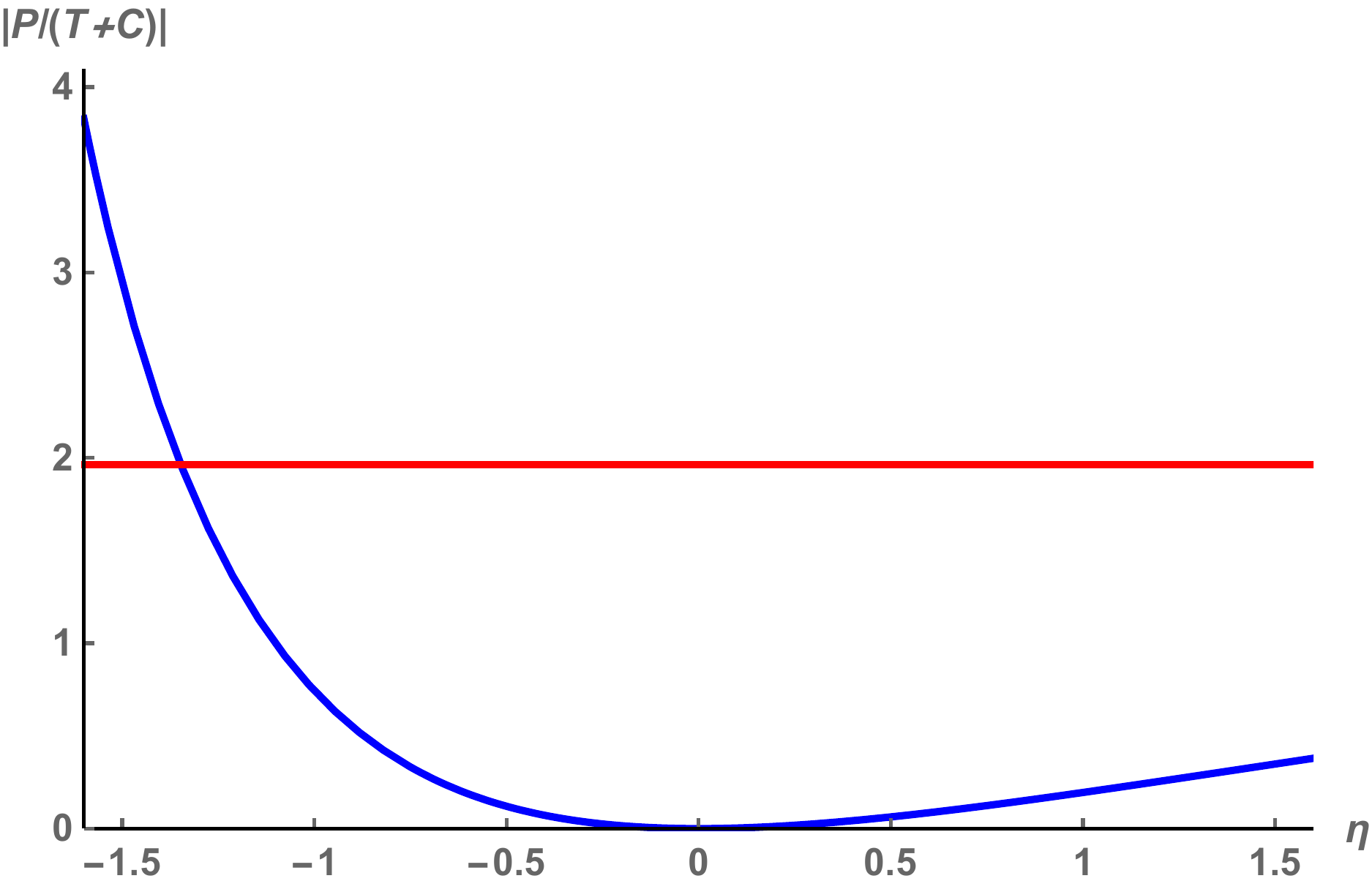}
\caption{Theoretical prediction for the ratio $|P/(T+C)|$ with $\eta$ varying from $-1.6\sim 1.6$ (blue) and $t=m_t^2$ (red). }
\label{fsi}
\end{figure}

The dependence of the ratio $|P/(T+C)|$ on the parameter $\eta$ is displayed in Fig.~\ref{fsi}.
It is found that $|P/(T+C)|$ is very sensitive to $\eta$.
With $\eta$ varying in the range between $-1.6$ and $1.6$, $|P/(T+C)|$ varies from zero to four.
The large $|P/(T+C)|$ in some area indicates that the large penguin is accessible in the non-perturbative QCD.
If we set $t=m_t^2$, $|P/(T+C)|$ is expected to be $1.96$.
It is consistent with the penguin amplitude extracted from $CP$ asymmetry in the $D^0\to \pi^+\pi^-$ mode.

\section{Summary}\label{con}

We studied $CP$ asymmetry in the $D\to \pi\pi$ system based on the isospin and topological analysis.
According to the new measurement of $CP$ asymmetry in the $D^0\to \pi^+\pi^-$ decay, we concluded that $CP$ asymmetry in the $D^0\to \pi^0\pi^0$ decay could reach to be $\mathcal{O}(10^{-3})$, which is available on Belle II in the future.
Besides, the large $CP$ asymmetry in the $D\to \pi\pi$ system might be explained by the $t$-channel final-state interaction.

\section{Acknowledgement}
We are grateful to Wei Shan for useful discussions.
This work was supported in part by the National Natural Science Foundation of China under Grants No. 12105099.


\end{document}